\documentstyle[prd,aps,eqsecnum,tighten]{revtex}
\begin{document}
\draft
\title{ K\"ahler Fermions on the Weitzenb\"ock space-time  . } 
\author{ I.~B.~Pestov\thanks{Electronic address:
pestov@thsun1.jinr.ru}}
\address{Bogoliubov Laboratory of Theoretical Physics, Joint
Institute for
Nuclear Research \\ Dubna, 141980, Russia}

\date{\today}
\maketitle
\begin{abstract}
 The K\"ahler-Dirac equation  is derived on the Weitzenb\"ock
space-time, which has a quadruplet of parallel vector fields as
the fundamental structure. A consistent system of equations for the
K\"ahler fields and parallel vector fields is obtained.


\end{abstract}

\section{Introduction} The notion of absolute parallelism or
teleparallelism was introduced by Einstein [1] when he tried  to
unify gravitation and electromagnetism.  A transcription of the Dirac
equation as a set of equations for an antisymmetric tensor field was
introduced by mathematician E. K\"ahler [2](see also [3]).This
K\"ahler-Dirac equation has been studied in connection with lattice
fermions [4],[5] and other remarkable properties [6]-[10]. As it was
suggested by Graf [6] the K\"ahler field might be more fundamental
than the Dirac spinor. This is an appealing idea because it conforms
to the Einstein methodology of associating all physical fields with
geometrical objects.

Here, the K\"ahler-Dirac equation on the Weitzenb\"ock space-time
[11] characterized by the vanishing curvature tensor (absolute
 parallelism) and by the torsion tensor formed of four parallel
 vector fields (parallel frames ) is studied.  Motivation for our
 consideration is the known fact that orthonormal frames are
 necessary for description of gravitational interactions of fermions
 in the framework of the usual spinor formalism.  To investigate the
 role of frames in the K\"ahler formalism, it is natural to consider
 the K\"ahler-Dirac equation on the Weitzenb\"ock space-time.  Our
 main result is the consistent system of equations (1),(2),(3),(4)
 for the K\"ahler fields and parallel vector fields with 16 degrees
 of freedom defined by the internal symmetry inherent in the
 K\"ahler-Dirac equation on the Weitzenb\"ock space-time.  We do not
 consider the physical interpretation of the equations derived
 because at that time it was absent for the K\"ahler field on a more
 familiar Riemann space-time, but the results of our consideration
 may be useful for solving this problem.

\section{K\"ahler-Dirac equation on the \protect{\\}Weitzenb\"ock space-time }

 The Weitzenb\"ock space-time admits a quadruplet of linearly
 independent parallel vector fields $h^{\mu}_i,$  defined by
 $$
				\nabla_{\nu} \,h^{\mu}_i = \partial_{\nu}
 h^{\mu}_i + \Gamma^{\mu}_{\nu\lambda} h^{\lambda}_i = 0.
$$
 Solving these equations, we find the non-symmetric connexion,

$$ \Gamma^{\lambda}_{\mu\nu} = h^{\lambda}_i
\partial_{\mu} h^i_{\nu}, $$
and the torsion tensor,
$$
S^{\lambda}_{\mu\nu}=\frac{1}{2}h^{\lambda}_i (\partial_{\mu}
h^i_{\nu} - \partial_{\nu} h^i_{\mu}). $$ Here $h^i_{\mu}
$  is also a quadruplet of parallel covector fields,  inverse
to $h^{\mu}_i. $  Thus, the coefficients $h^{\mu}_i $ or $h^i_{\mu}
,$ are  16 functions and must satisfy the relations$$h^{\mu}_i
h^i_{\nu}= \delta^{\mu}_{\nu}, \quad h^{\mu}_jh^i_{\mu}=\delta^i_j,
$$ $$ \eta_{ij}h^i_{\mu} h^j_{\nu} = g_{\mu\nu}, \quad
\eta^{ij}h_i^{\mu} h_j^{\nu} = g^{\mu\nu},$$ where $\eta_{ij} =
\eta^{ij} = \rm diag(+,-,-,-).$

The covariant antisymmetric
tensor field $f_{{\mu}_1 \cdots {\mu}_p} \quad (p=0,1,2,3,4)$  is
 called the p-form.  If $$F = (f, f_{\mu}, f_{\mu\nu},
 f_{\mu\nu\lambda}, f_{\mu\nu\lambda\sigma})$$ is an
 inhomogeneous form,(i.e. K\"ahler field) the generalized curl
 operator $D_e$ is defined by $$D_e F = (0, D_\mu f, 2D_{[\mu}
 f_{\nu]}, 3D_{[\mu} f_{\nu\lambda]}, 4 D_{[\mu}
 f_{\nu\lambda\sigma]}) ,$$ where $D_{\mu} = \nabla_{\mu} + S_{\mu},$
 the covector $S_{\mu}$ is equal to the contraction of the torsion
tensor $S_{\mu} = S^{\tau}_{\mu\tau}.$ Square brackets denote
alternation.  For the operator $D_i$ of generalized divergence we
 have the following definition $$D_i F = (- D^{\tau} f_{\tau} , -
 D^{\tau} f_{\tau\mu}, - D^{\tau} f_{\tau\mu\nu}, - D^{\tau}
 f_{\tau\mu\nu\lambda}, 0).$$ The K\"ahler-Dirac equation on the
 Weitzenb\"ock space-time  is of the form \begin{equation} D F =
 \frac{mc}{\hbar} F, \end{equation} where $D = D_i+ D_e.$

 Similar to
 the operators $D_i$ and $D_e$ one can introduce the operators $Q_i$
 and $Q_e $ defined by the vector field $u^{\mu}$ as
 follows $$Q_e F = (0, u_{\mu} f, 2u_{[\mu} f_{\nu]},
 3u_{[\mu} f_{\nu\lambda]}, 4 u_{[\mu} f_{\nu\lambda\sigma]}) ,$$

 $$Q_i F = (- u^{\tau} f_{\tau} , - u^{\tau} f_{\tau\mu}, - u^{\tau}
 f_{\tau\mu\nu}, - u^{\tau} f_{\tau\mu\nu\lambda}, 0).$$ If in
 additional to these operators, we introduce  a numerical operator
 $\Lambda$ such that $$ \Lambda F = (f, -f_{\mu}, f_{\mu\nu},
 -f_{\mu\nu\lambda}, f_{\mu\nu\lambda\sigma}),$$ then it can be shown
 that the operator $Q = (Q_i - Q_e) \Lambda$ commutes with the
 operator $D$ under the condition $\nabla_{\mu} u^{\nu} = 0.$
 Thus, the operator $Q$ acts in the space of the solutions
 of  equation (1) if $u^{\mu} = u^i h^{\mu}_i,$ where
 $u^i $ are constants.

 This internal symmetry of equation (1) immediately
 gives the tensor $$ J^{\mu\nu} = \sum\nolimits_{p=0}^{4}
 \frac{(-1)^{p+1}}{p!}( \frac{1}{2}g^{\mu\nu} \bar f^{{\tau}_1 \cdots
 {\tau}_p}   f_{{\tau}_1 \cdots {\tau}_p}+ \bar f^{{\mu}{\tau}_1
 \cdots {\tau}_p} f^{\nu}_{.{\tau}_1 \cdots {\tau}_p} $$ $$ -
 \bar f^{{\mu\nu} {\tau}_1 \cdots {\tau}_p}  f_{{\tau}_1 \cdots
  {\tau}_p}) + c.c., $$ which on the solutions of  equation (1)
 satisfies the equation \begin{equation} (\nabla_{\tau} + 2S_{\tau})
J^{\tau\mu} =  0. \end{equation}
To prove this, let
$$(F,H) = \sum\nolimits_{p=0}^{4} \frac{1}{p!}( \bar f^{{\tau}_1
 \cdots {\tau}_p} f_{{\tau}_1 \cdots {\tau}_p}).  $$
 After some calculations it can be shown that
 $$(DF + m F, H) - (F, DH + m H) = $$
 $$(\nabla_{\tau}+  2S_{\tau})  R^{\tau} ,$$
 where  $R^{\tau} = (F,H)^{\tau}$ and
 $$ (F,H)^{\tau} = \sum\nolimits_{p=0}^{4} \frac{1}{p!}( \bar
 f_{{\mu}_1 \cdots {\mu}_p} h^{\tau{\mu}_1 \cdots {\mu}_p} -
 \bar f^{\tau{\mu}_1 \cdots {\mu}_p}h_{{\mu}_1 \cdots {\mu}_p}). $$
 From this equality it follows that if $F$ and $H$ are solutions to
 equation (1), then the vector $R^{\tau} =(F,H)^{\tau} $
 satisfies the equation
 $$ ( \nabla_{\tau} + 2S_{\tau} )R^{\tau} = 0.$$
 Substitution $H= QF$ gives
 $$(F,QF)^{\tau} = u_{\mu} J^{\tau\mu},$$ which proves  the
 statement.

 \section{Equation for parallel vector fields }

Since the tensor $J^{\tau\mu}$ has 16 components, one can write
 the consistent system of equations for 16 functions $h^{\mu}_i$
 \begin{equation}
			 E^{\mu\nu} = l J^{\mu\nu}
 \end{equation}
provided that from the components of $h^{\mu}_i$ and their
 first and second derivatives one can construct the tensor
 $E^{\mu\nu}$ which satisfies the equation $$ (\nabla_{\tau} +
 2 S_{\tau}) E^{\tau\mu} = 0 $$ identically.   It is remarkable that
a tensor like that really exists.

 Let
    $$S^{\mu\nu\lambda} = g^{\mu\tau} g^{\nu\sigma}
 S^{\lambda}_{\tau\sigma}.$$ From the commutator of covariant
 derivatives $$\nabla_{\tau} \nabla_{\sigma} S^{\mu\nu\lambda} -
 \nabla_{\sigma} \nabla_{\tau} S^{\mu\nu\lambda} = - 2
 S^{\beta}_{\tau\sigma} \nabla_{\beta} S^{\mu\nu\lambda} $$
 it follows that
 $$
	 \nabla_{\mu}( \nabla_{\nu} S^{\mu\nu\lambda} +
 S^{\mu}_{\tau\sigma} S^{\tau\sigma\lambda}) =
 S^{\tau\sigma\lambda} \nabla_{\mu} S^{\mu}_{\tau\sigma}
 .$$
The right hand side of this equality admits following representation.
The identity for general linear connexion [12]
 $$R_{[\nu\tau\sigma]}{.}^{\mu} =2 \nabla_{[\nu} S^{\mu}_{\tau\sigma]}
 -4 S^{\beta}_{[\nu\tau} S^{\mu}_{\sigma]\beta}$$
  in our case $(R_{\nu\tau\sigma}{.}^{\mu}=0 )$ gives
 $$\nabla_{\nu} S^{\mu}_{\tau\sigma} +\nabla_{\tau}
 S^{\mu}_{\sigma\nu}+ \nabla_{\sigma} S^{\mu}_{\nu\tau} = $$ $$
2S^{\beta}_{\nu\tau} S^{\mu}_{\sigma\beta} + 2S^{\beta}_{\tau\sigma}
 S^{\mu}_{\nu\beta}+ 2S^{\beta}_{\sigma\nu} S^{\mu}_{\tau\beta}  . $$
 From this one derives by contraction
 $$\nabla_{\mu}S^{\mu}_{\tau\sigma} = - \nabla_{\tau}S_{\sigma} +
 \nabla_{\sigma}S_{\tau} - 2S_{\beta} S^{\beta}_{\tau\sigma}.$$
 Thus,
 $$
 S^{\tau\sigma\lambda}  \nabla_{\mu} S^{\mu}_{\tau\sigma}
 = -\nabla_{\mu} (2 S_{\nu} S^{\mu\nu\lambda}) -
 2S_{\mu}( \nabla_{\nu}S^{\mu\nu\lambda} +
 S^{\mu}_{\tau\sigma} S^{\tau\sigma\lambda}).
 $$
 After substitution in to the starting equality we get
$$
	 \nabla_{\mu}( \nabla_{\nu} S^{\mu\nu\lambda} +
 S^{\mu}_{\tau\sigma} S^{\tau\sigma\lambda} + 2S_{\nu}
 S^{\mu\nu\lambda}) =$$ $$ -
	2S_{\mu}( \nabla_{\nu} S^{\mu\nu\lambda} +
 S^{\mu}_{\tau\sigma} S^{\tau\sigma\lambda} + 2S_{\nu}
 S^{\mu\nu\lambda})
.$$
and hence the tensor
\begin{equation}E^{\mu\lambda} =\nabla_{\nu} S^{\mu\nu\lambda} +
S^{\mu}_{\tau\sigma} S^{\tau\sigma\lambda} + 2S_{\nu}S^{\mu\nu\lambda}
\end{equation}
 satisfies  equation (2)  identically.

Thus, we have proved that the system of
equations (1),(2),(3),(4)  is consistent.

 Since the tensors $J^{\mu\nu} and
\quad E^{\mu\nu}$ have dimensions $cm^{-3}$ Ánd respectively
 $cm^{-2},$ then the constant $l$ in  equation (3) has the dimension
 of length.  It is natural to suppose that $l$ is equal to the Planck
 length. It should be noted also that the 00-component of the tensor
 $J^{\mu\nu}$ is positive definite.

 Thus, the connection is established between the theory of  K\"ahler
 fermions and the spaces with teleparallelism, expressed by the
 system of equations (1),(2),(3),(4). \begin{flushleft} {\bf
 REFERENCES} \end{flushleft}

\begin{enumerate}
\item  Einstein, A. (1930).{\sl  Sitzungsber. preuss. Acad. Wiss.,
phys.-math.  Kl.}, 401. \item K\"ahler, E. (1962). {\sl Rend.
Mat.},{\bf (3-4) 21}, 425.   \item  Ivanenko, D.,  and
Landau, L. (1928). {\sl Zeit. fur Phys.,}  {\bf 48}, 340.\item
Becher, P., Joos, H. (1982). {\sl Zeit. fur Phys.,}, {\bf  C15},
343. \item  Gockeler, M. (1984). {\sl Phys.  Lett.}, {\bf B142},
197. \item  Graf, W. (1978). {\sl  Ann.  Inst.  Henri Poincare},
{\bf A29,} 85.   \item Banks, I.M., Dothan, V., Horn,
D. (1982). {\sl Phys.  Lett.,} {\bf B117}, 413.  \item
Benn, I.M., Tucker, R.W. (1982). {\sl Phys.  Lett.,} {\bf B119,}
348.   \item Holdom, B. (1984). {\sl Nucl. Phys.,} {\bf B233,}
413.   \item  Bullinaria, J.A. (1985). {\sl Ann.  of Phys.,}
{\bf159,} 272. \item Weitzenb\"ock, R. (1923). {\sl
Invariantentheorie}{\bf XIII, 7}. (Noordhoff, Groningen). \item
Schouten, J.A. (1954). {\sl Ricci-Calculus}. (Berlin).
\end{enumerate}

\end{document}